\documentclass[8pt]{iopart}
\usepackage{iopams} 
\usepackage[latin1]{inputenc}
\usepackage{graphicx}
\usepackage{url}

\begin{document} 
 
\title[Graphene plasmonic antenna -- graphene surface plasmon]{Theoretical investigation of the spontaneous emission on graphene plasmonic antenna in THz regime}
\author{Mauro Cuevas} 
\address{Consejo Nacional de
Investigaciones Cient\'ificas y T\'ecnicas (CONICET) and Facultad de Ingenier\'ia y Tecnolog\'ia Inform\'atica, Universidad de Belgrano,
Villanueva 1324, C1426BMJ, Buenos Aires, Argentina}
\address{Grupo de Electromagnetismo Aplicado, Departamento de F\'isica, FCEN, Universidad de Buenos Aires,  Ciudad Universitaria,
Pabell\'on I, C1428EHA, Buenos Aires, Argentina}
\ead{cuevas@df.uba.ar}

\begin{abstract} 
The present work deals with a theoretical research on the emission and radiation properties of a dipole emitter  source close to a dimer graphene plasmonic antenna.  Modification of the radiation and the quantum efficiencies resulting from varying the position of the emitter and the orientation of its dipole moment are calculated by using a rigorous electromagnetic method based on Green's second identity. 
Large enhancements in the emission and the radiation of the emitter occur due to the coupling with the  antenna surface plasmons in the spectral region from $\approx 4$THz to $\approx 15$THz.    
Our results show that the radiation efficiency can be enhanced by four orders of magnitude and that the quantum efficiency reaches values close to $0.8$  when the emission frequency coincides with one of the resonant  dipolar frequencies.  
On the other hand, these quantities can be  reduced in a great measure at a specific frequency for a given emitter location.  We present calculations of the near--field distribution and the far field intensity which reveal the role of the plasmonic antenna resonance in the emitter enhanced radiation. We show that the spectral region where the radiation is enhanced can be chosen over
a wide range by varying the chemical potential of graphene from $0.2$eV to $1$eV. 
\end{abstract} 



\maketitle
\ioptwocol

\section{Introduction} 

The coherent interplay between an individual optical emitter and the electromagnetic fields scattered back to the emitter's site by the environment boundaries acts to drive the emitter and as a consequence the light emission can be largely altered. For instance, coupling the optical emitter to confined electromagnetic modes, such as guided modes or surface plasmons (SPs), is possible to enhance  the spontaneous emission rate several orders of magnitude 
relative to the case in which the same emitter is localized in an unbounded medium.  
This property, known as Purcell effect \cite{purcell,maier_nature}, has been used to improve the efficiency of single photon sources \cite{SPS} as well as to reduce the lasing threshold in plasmonic lasers \cite{laser,laser1}. 

Based on the same physical effect, plasmonic antennas are subwavelength architectures  capable to confine the electromagnetic field in a reduced region of the space by localized surface plasmons (LSPs) excitation, leading to an enhancement in the radiative and non radiative emission of an emitter placed in that region \cite{SPA,VH,alu2,christensen,skigin,sanchezgil2}. 
One of the major challenges for most of applications, such as fluorescence or surface enhanced Raman 
spectroscopy (SERS) platforms \cite{novotny,rogobete1,lalanne}, is to reach an efficient radiative 
outcoupling of LSPs into photons, which allows an increase of the  radiative  emission rate relative to the 
non--radiative  emission rate. This property has been obtained for some geometries, such as dipole or bowtie antennas \cite{rogobete2,martin,ren}, stand out for creating a high density of radiative states on their gap region. More recently, its has been demonstrated that a high radiation efficiency together with a  directionality improvement take place on metallic nanopatch antenna platforms \cite{NP,NP1}.

Apart from noble metals, the plasmonic materials most frequently used for plasmonic antenna applications,
recent advances have created other plasmonic materials with lower losses and greater confinement of the electromagnetic field, such as metal--alloys, heavily doped wide--band semiconductors, and graphene \cite{shalaev}. The  electronic linear band structure of graphene makes a plasmon mass depending on the Fermi--level position and consequently electrically (or magnetically) tunable SPs are supported by graphene from microwaves to the mid--infrared regimes \cite{jablan,Xia}. Several alternative structures for confining of the  incident  beam in the realm of graphene plasmonic have been studied  in THz regime \cite{cuevasATR,moradi,farmani1,farmani2,farmani3,shadrivov,nikitin,Pashaeiad}. 
%
%
%
%
In particular, the high SP confinement on a graphene monolayer leads to two main properties: a small SP wavelength and an improvement in the electromagnetic energy density. The former deals with the possibility to build more smaller plasmonic constituent elements and the second is related with the large enhancement of the decay rate of an emitter via the Purcell effect \cite{ribons,grafeno4,cuevas0}. These features positioned the graphene as a promising platform to the development of controllable plasmon devices \cite{challenges,farmani,lai}, in particular of a new generation of   antennas  from microwaves to the mid--infrared regimes \cite{antenas1} which comprises a crossover between electronic and optics. 
In this way, graphene antennas have found applications as sensors, \textit{i.e.}, to capture an impinging wave in order to feed oscillating dipoles \cite{alu}, or as emitters, where the enhanced density of radiating states enables an excitation of oscillating dipoles \cite{filter}.  In fact, graphene antennas have been proposed as reconfigurable transceivers in the THz domain \cite{gomez_diaz}.

This paper deals with the study about the control of the radiative emission of a single emitter placed in the environment of a graphene plasmonic antenna. We focus on the  dipole geometry in which the antenna consists  of two identical  elements facing each other (nearly in contact). We consider each of these constituent elements consisting of a wire cylinder (arbitrary section) wrapped   with a graphene sheet. The calculation of relevant magnitudes is carried out by using the Green function surface integral method (GSIM) \cite{sanchezgil,valencia,GSIM} which enables to solve the scattering problem for structures with a complex shape. The GSIM  has been used by us to deal with the electromagnetic scattering of an optical emitter inside a graphene coated wire of arbitrary section \cite{cuevas3}.

We have considered the graphene as an  infinitesimally thin, two--sided layer with a frequency--dependent surface conductivity given by the Kubo formula \cite{falko}. This approach is particularly appropriate and, as it has been shown in \cite{merano}, it matches the results in  remarkable optical experiments.

This paper is organized as follows. First, in Section \ref{teoria} we develop the GSIM formalism providing  expressions for the electromagnetic field scattered by a line dipole  source placed, with an arbitrary orientation, near a  graphene plasmonic antenna. 
 This field is expressed in terms of two unknown source functions evaluated on the graphene layer, one related to the field exterior to the antenna and the other related to its normal derivative. By using GSIM, in Section \ref{resultados} we calculate the radiative and the quantum efficiencies for a graphene dimer antenna.
Even though the above formalism has been developed considering two scatter elements (the two graphene wire composing the antenna), for a better understanding,  in a first step we deal with the single graphene wire problem. We then include a second graphene wire and investigate the performance of the dimer antenna formed. Finally,  concluding remarks are provided in Section \ref{conclusiones}. 
The Gaussian system of units is used
 and an $\mbox{exp}(-i\, \omega\, t)$ time--dependence is implicit throughout the paper, with $\omega$ as the angular frequency, $t$ as the time, and $i=\sqrt{-1}$. The symbols Re and Im are respectively used for denoting the real and imaginary parts of a complex quantity.

\section{Theory} \label{teoria} 

\subsection{Surface integral equations of the electromagnetic field scattered by a dipole emitter}

We consider the scattering problem of a line dipole source in the proximity of two  graphene coated wire cylinders (figure 1). We assume that the cylinders and the dipole line axis lie along the $\hat{z}$ axis.   The current density of the electric dipole  is
\begin{eqnarray}\label{corriente}
\vec{j_s}(\vec{r})=-i \omega \vec{p} \, \delta(\vec{r}-\vec{r_s}).
\end{eqnarray} 
\begin{figure}
\centering
\resizebox{0.40\textwidth}{!}
{\includegraphics{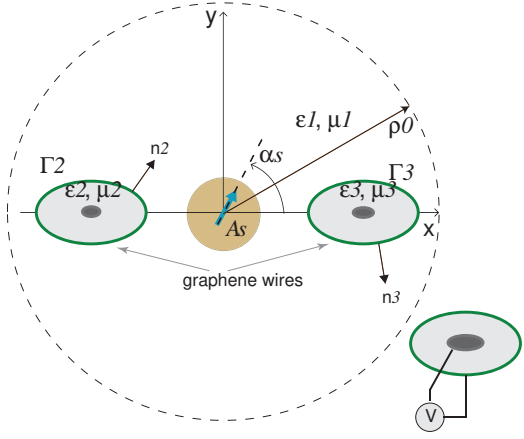}}
\caption{\label{fig:epsart} Schematic illustration of the dimer micro antenna. The inset shows a contact  rod  embedded inside the graphene coated dielectric cylinder needed to control the chemical potential with a voltage $V$ application. 
}\label{sistema}
\end{figure}

The cross section of the wires are defined by planar curves described by  vector valued functions $\Gamma_2(t)=x_2(t)\hat{x}+y_2(t)\hat{y}$ and $\Gamma_3(t)=x_3(t)\hat{x}+y_3(t)\hat{y}$ and the wire substrates are  characterized by constitutive parameters $\varepsilon_2,\,\mu_2$ and  $\varepsilon_3,\,\mu_3$. The wires are  embedded in a transparent medium with constitutive parameters  $\varepsilon_1,\,\mu_1$. 
When the line source with a dipole moment $\vec{p}=p [\cos \alpha \, \hat{x}+ \sin \alpha\,\hat{y}]$ is placed in medium 1 ($\alpha$ is the angle between the dipole moment and the $\hat{x}$ axis), the magnetic field is along the $\hat{z}$ axis ($\vec{H}(\vec{r})=\varphi(\vec{r}) \, \hat{z}$). 
The wave equation for the magnetic field has the form
%
%
%
\begin{eqnarray}\label{A}
\nabla^2 \varphi_j(\vec{r})+k_j^2 \varphi_j(\vec{r}) = g_j(\vec{r},\vec{r}_s),
\end{eqnarray} 
where subscripts $j=1,\,2,\,3$ is used to denote the wire substrates (medium 2 and medium 3) and the exterior region (medium 1)  to boundary wires, respectively, $k_j=k_0 \sqrt{\varepsilon_j\,\mu_j}$, 
$k_0=\omega/c$ is the modulus of the photon wave vector in vacuum, $\omega$ is the angular frequency, $c$ is the vacuum speed of light, $g_1(\vec{r},\vec{r}_s)=-4\,\pi\, i k_0 \,\vec{p} \times\,\nabla \delta(\vec{r}-\vec{r}_s)$, $g_2(\vec{r},\vec{r}_s)=g_3(\vec{r},\vec{r}_s)=0$ and $\vec{r}_s$ denotes the position of the line source.  
To solve Eq. (\ref{A}), we transform it into a boundary integral equation using the GSIM as explained in \cite{cuevas3}. Using Eq. (\ref{A}) in the exterior region, the magnetic field $\varphi(\vec{r})$ can be writen as 
\begin{equation}\label{phi1}
\begin{array}{ll}
\varphi_{inc}(\vec{r})\\
+\sum_{j=2}^3 \frac{1}{4\pi}
\int_{\Gamma_j}\,\left[\frac{\partial\,G_1(\vec{r},\vec{r_j})}{\partial n_j}\varphi_1(\vec{r_j})-G_1(\vec{r},\vec{r_j}) \frac{\partial\,\varphi_1(\vec{r_j})}{\partial n_j}\right]\,ds_j\\
& \\
=\left\{
\begin{array}{ll}
\varphi_1(\vec{r})\,\,\,\mbox{if $\vec{r}$ is inside the exterior region 1}\\
0 \,\,\,\mbox{if $\vec{r}$ is outside the exterior region 1}
\end{array}
\right.
\end{array}
\end{equation} 
where 
$ds_j$ is   the  arc element of $\Gamma_j$ ($j=2,\,3$), the derivative   $\frac{\partial}{\partial n_j}$  along the normal to the interface  at  $\vec{r_j}$ is directed from the medium $j$ ($j=2,\,3$) to the medium 1, 
and $G_1(\vec{r},\vec{r_j})$ is the Green function of Eq. (\ref{A}) in the exterior region (medium 1)
\begin{eqnarray}\label{funcion de green}
G_1(\vec{r},\vec{r_j})=i \pi H_0^{(1)}(|\vec{r}-\vec{r_j}|), 
\end{eqnarray}
where $H_0^{(1)}$ is the 0th Hankel functions of the first kind, and 
\begin{eqnarray}\label{inc}
\varphi_{inc}(\vec{r})=i k_0 \hat{z}\, (\vec{p} \times \nabla)  G_1(\vec{r},\vec{r}_s).
\end{eqnarray}
Similarly, inside  the wires (regions 2 and 3) the field take the form 
\begin{equation}\label{phij}
\begin{array}{ll}
-\frac{1}{4\pi}\int_{\Gamma_j}\,\left[\frac{\partial\,G_j(\vec{r},\vec{r_j})}{\partial n_j}\varphi_j(\vec{r_j})-G_j(\vec{r},\vec{r_j}) \frac{\partial\,\varphi_j(\vec{r_j})}{\partial n_j}\right]\,ds_j \\
&\\
=\left\{
\begin{array}{ll}
\varphi_j(\vec{r})\,\,\,\mbox{if $\vec{r}$ is inside the interior region $j$},\\
0 \,\,\,\mbox{if $\vec{r}$ is outside the interior region $j$}
\end{array}
\right.
\end{array}
\end{equation} 
where $G_j(\vec{r},\vec{r_j})$ is the Green function in the interior region $j=2,\,3$ to the wires. From Eqs. (\ref{phi1}) and (\ref{phij}), the total field in regions 1, 2 and 3 are completely determined by the boundary values of the field and its normal derivative. By allowing the point of observation $\vec{r}$ to approach the surface in Eqs. (\ref{phi1}) and (\ref{phij}), we obtain a system of coupled integral equations with unknown functions and normal derivatives at the $\Gamma_j$ ($j=2,\,3$) boundaries.      
The electromagnetic boundary conditions at $\Gamma_j$ ($j=2,\,3$),
\begin{eqnarray}\label{cc1}
\frac{1}{\varepsilon_j}\frac{\partial \varphi_j}{\partial n_j}=\frac{1}{\varepsilon_1}\frac{\partial \varphi_1}{\partial n_j},
\end{eqnarray} 
and 
\begin{eqnarray}\label{cc2}
\varphi_1-\varphi_j= 
\frac{4 \pi \sigma}{c k_0 \varepsilon_1}i \frac{\partial \varphi_j }{\partial   n_1}, 
\end{eqnarray} 
%
provide two additional relationships between the fields and their normal derivatives at the boundary of the wires,  allowing us to express $\varphi_j$, $\partial \varphi_j / \partial n_j$ ($j=2,\,3$) in terms of  $\varphi_1$ and $\partial \varphi_1 / \partial n_j$. 
By evaluating the first of Eq. (\ref{phi1}) and the second of Eq. (\ref{phij}) at the boundaries $\Gamma_j$ and using 
the continuity conditions across them, we obtain a set of coupled integral equations:
\begin{equation}\label{s1}
\begin{array}{ll}
\varphi_1(\vec{r_l}')=\varphi_{inc}(\vec{r_l}')\\
& \\
+\sum_{j=2}^3 \frac{1}{4\pi}\\
\times
\int_{\Gamma_j}\,\left[\frac{\partial\,G_1(\vec{r_l}',\vec{r_j})}{\partial n_j}\varphi_1(\vec{r_j})-G_1(\vec{r_l}',\vec{r_j}) \frac{\partial\,\varphi_1(\vec{r_j})}{\partial n_j}\right]\,ds_j,
\end{array}
\end{equation} 
where $l=2,\,3$ indicates the region in which the vector $\vec{r}$ is taken, and 
\begin{equation}\label{s2}
\begin{array}{ll}
0=
-\frac{1}{4\pi} \int_{\Gamma_j}\,[\frac{\partial\,G_j(\vec{r_l}',\vec{r_j})}{\partial n_j}\varphi_1(\vec{r_j})-\\
& \\
\left\{ \frac{\varepsilon_j}{\varepsilon_1}G_j(\vec{r_l}',\vec{r_j}) + \frac{4 \pi \sigma}{c k_0 \varepsilon_j} i \frac{\partial\,G_j(\vec{r_l}',\vec{r_j})}{\partial n_j}\right\} \frac{\partial\,\varphi_1(\vec{r_j})}{\partial n_j} ]\,ds_j
\end{array}
\end{equation} 
where $l,j=2,\,3$.

It is worth noting that the four Eqs. (\ref{s1}) and (\ref{s2})  are reduced to the set of two equations  obtained  in \cite{cuevas3} for the single wire case, \textit{i.e.}, when the system is composed of only one graphene wire element whose cross section is defined by a vector valued function $\Gamma_2$,    by fixing $j=l=2$,

\begin{equation}\label{s1wire}
\begin{array}{ll}
\varphi_1(\vec{r_2}')=\varphi_{inc}(\vec{r_2}')\\
& \\
+\frac{1}{4\pi}
\int_{\Gamma_2}\,\left[\frac{\partial\,G_1(\vec{r_2}',\vec{r_2})}{\partial n_2}\varphi_1(\vec{r_2})-G_1(\vec{r_2}',\vec{r_2}) \frac{\partial\,\varphi_1(\vec{r_2})}{\partial n_2}\right]\,ds_2,\\
& \\
0=
 -\frac{1}{4\pi}
\int_{\Gamma_2}\,[\frac{\partial\,G_2(\vec{r_2}',\vec{r_2})}{\partial n_2}\varphi_1(\vec{r_2})-\\
& \\
\left\{ \frac{\varepsilon_2}{\varepsilon_1}G_2(\vec{r_2}',\vec{r_2}) + \frac{4 \pi \sigma}{c k_0 \varepsilon_2} i \frac{\partial\,G_2(\vec{r_2}',\vec{r_2})}{\partial n_2}\right\} \frac{\partial\,\varphi_1(\vec{r_2})}{\partial n_2} ]\,ds_2.
\end{array}
\end{equation} 

The coupled integral equations (\ref{s1})--(\ref{s2}), or Eq. (\ref{s1wire}) in case of the single wire system, are converted into matrix equations which are solved numerically  (see \cite{cuevas3} and references therein). Once the functions $\varphi_1(\vec{r_j})$ and $\frac{\partial\,\varphi_1(\vec{r_j})}{\partial n_j}$ 
are determined, the scattered field, given by the first of Eqs. (\ref{phi1}) and (\ref{phij}) can be calculated at every point in the exterior and  interior  regions.     

\subsection{Emitted and radiated powers}

The time--averaged  power $P$ emitted by the dipolar line source can be calculated from the integral of the normal component of the complex Poynting vector flux through an imaginary cylinder of length $L$ and surface $A_s$ that encloses the dipole (see Figure 1)
%
%
%
\begin{eqnarray}\label{P}
P= -\frac{1}{2}L\int_{A_s} \Re \left\{\vec{j}_s^{*} \cdot \vec{E}\right\} da,
\end{eqnarray}
where $A_s$ encloses the source, $da=\rho_s d\phi_s d\rho_s$ and $\vec{j}_s$ represents the source density current.  Introducing the value of the current in Eq. (\ref{corriente}), we obtain 
\begin{eqnarray}\label{P2}
P=\frac{\omega L}{2}\mbox{Im}\left\{\vec{p}^* \cdot \vec{E}(\vec{r_s}) \right\}
\end{eqnarray}  
where the field $\vec{E}$ is evaluated at the dipole position $\vec{r_s}$. For an electric dipole we have
\begin{eqnarray}\label{E0Es}
\vec{E}(\vec{r})=\vec{E_{inc}}(\vec{r})+\vec{E}_s(\vec{r})
\end{eqnarray}  
%
where $\vec{E}_{inc}$ and $\vec{E}_s$ are the primary dipole field and the scattered field, respectively.
Inserting Eq. (\ref{E0Es}) into Eq. (\ref{P2}),   we obtain   the emitted power normalized with respect to the rate in absence of the  graphene wire
%
%
%
%
\begin{equation}\label{P1}
\begin{array}{ll}
\gamma=\frac{P}{P_{inc}}=\nonumber\\
& \\
1+\frac{\Im\left\{\vec{p} \cdot \vec{E}_s(\vec{r_s})\right\}}{\Im\left\{\vec{p} \cdot \vec{E}_{inc}(\vec{r_s})\right\}}=
1+4 \frac{ \Im\left\{\vec{p} \cdot \vec{E}_s(\vec{r_s})\right\}}{\pi k_0^3 p^2 c},
\end{array}
\end{equation}
where $P_{inc}=\frac{\pi k_0^3 p^2 c L}{4}$ is the total power radiated by an electric dipole in the unbounded medium 1 \cite{cuevas3,cuevas3bis}. By using the Ampere--Maxwell equation the relation between the components of
the electric field  and the z--component of the magnetic field is derived, $\vec{E}_j = -\frac{1}{ i k_0 \varepsilon_j} \nabla \times \hat{z} \varphi_j$ ($j=1,\,2,\,3$), and the components of the electric field scattered back at the dipole position $\vec{r_s}$ are calculated as 
\begin{equation}\label{Ex}
\begin{array}{ll}
E_{s,x}(\vec{r_s})=-\frac{1}{ik_0\varepsilon_1}
& \\
\times \sum_{j=2}^3 \int_{\Gamma_j} \left[ \frac{\partial^2 G_j(\vec{r_s},\vec{r_j})}{\partial y \partial n_j}\varphi_1(\vec{r_j})-\frac{\partial G_j(\vec{r_s},\vec{r_j})}{\partial y} \frac{\partial \varphi_1(\vec{r_j})}{\partial n_j}\right] ds_j,  \\
& \\
E_{s,y}(\vec{r_s})=\frac{1}{ik_0\varepsilon_1}
& \\
\times \sum_{j=2}^3 \int_{\Gamma_j} \left[ \frac{\partial^2 G_j(\vec{r_s},\vec{r_j})}{\partial x \partial n_j}\varphi_1(\vec{r_j})-\frac{\partial G_j(\vec{r_s},\vec{r_j})}{\partial x} \frac{\partial \varphi_1(\vec{r_j})}{\partial n_j}\right] ds_j.  \\
\end{array}
\end{equation}

Similarly, the time--averaged radiative power  can be  evaluated by calculating the complex Poynting vector flux through an imaginary cylinder of length $L$ and radius $\rho_0$ that encloses the system  (see Figure 1)
\begin{eqnarray}\label{Pr}
P_{s}=\nonumber\\
\frac{ \rho_0\,L\,c^2}{8 \pi \omega \varepsilon_1} \int_0^{2 \pi} \mbox{Re} \left\{ -i\, \left[\varphi_1(\rho_0,\phi)\right]^\ast \frac{\partial \varphi_1(\rho_0,\phi)}{\partial \rho} \right\} d\phi.
\end{eqnarray}
In the far--field region the calculation of the scattered fields given by the first of the Eq. (\ref{phi1}) can be greatly simplified using the asymptotic expansion of the Hankel function for large argument \cite{abramowitz}. After some algebraic manipulation, we obtain 
\begin{eqnarray}\label{asintotico1}
\varphi_1(\vec{r})=\pi k_0 k_1 p \sin(\phi-\alpha) \sqrt{\frac{2}{\pi k_1 \rho}} e^{i(k_1\rho-\frac{3\pi}{4})} f_{ang}(\phi)+\nonumber\\
i\frac{e^{i(k_1\rho-\frac{\pi}{4})}}{\sqrt{8 \pi k_1 \rho}} \sum_{j=2}^3 F_{ang}^{(j)}(\phi),
\end{eqnarray}
\begin{eqnarray}\label{asintotico2}
\frac{\partial \varphi_1(\vec{r}) }{\partial \rho}=i k_1 \varphi_1(\vec{r})
\end{eqnarray}
where the angular factors are given by
\begin{equation}
\begin{array}{ll}
f_{ang}(\phi)=e^{-ik_1(\cos \phi\, x_s+\sin \phi\,y_s)},\\
& \\
F_{ang}^{(j)}(\phi)=\\
\int_{\Gamma_j(t)} [i k_1 (-y'_j(t) \cos \phi\, + x'_j(t) \sin \phi) \varphi(\vec{r_j}(t))\\
-\frac{\partial \varphi_1(\vec{r_j}(t))}{\partial n_j}] dt.
\end{array}
\end{equation}
When Eqs. (\ref{asintotico1}) and (\ref{asintotico2}) are substituted into Eq. (\ref{Pr}), the normalized radiative decay rate 
is written as
\begin{equation}\label{Pr2}
\begin{array}{ll}
\gamma_s=\frac{P_s}{P_{inc}}= 1+\\
\frac{c}{16 \pi^3 k_0^4\varepsilon_1 p^2} \int_0^{2\pi} [|F_{ang}^{(2)}(\phi)|^2+|F_{ang}^{(3)}(\phi)|^2\\
& \\
-8\pi p \sin (\phi-\alpha) k_1 k_0 \varepsilon_1 \mbox{Re}\{f_{ang}(\phi) [F_{ang}^{(2)}(\phi)\\
& \\
+F_{ang}^{(3)}(\phi)]^*\}+2 \mbox{Re}\{F_{ang}^{(2)}(\phi)F_{ang}^{(3)}(\phi)^*\} ] d\phi.
\end{array}
\end{equation}
It worth noting that by taking $F_{ang}^{(3)}=0$ in Eq. (\ref{Pr2}) we obtain the power radiated by a system conformed by only one graphene wire close to a dipole emitter (single wire system).

\section{Results}\label{resultados}

In this section we apply the formalism sketched in  previous sections to calculate the normalized 
radiative  $\gamma_s=P_{s}/P_{inc}$ decay rate and the radiative quantum efficiency $\eta=\gamma_s/\gamma$ that specify the fraction of energy emitted as radiation.  We assume that  the curvature radius of the wires is sufficiently large as to describe their optical properties as those of a wire characterized by the same surface conductivity as planar graphene.

Taking into account the description of experimental results by the zero--thickness interface model \cite{merano}, we consider the  graphene layer as an infinitesimally thin, local and isotropic two--sided layer with frequency--dependent surface conductivity $\sigma(\omega)$ given by the Kubo formula \cite{falko,milkhailov}, which can be read as  $\sigma= \sigma^{intra}+\sigma^{inter}$, with the intraband and interband contributions being
\begin{equation} \label{intra}
\sigma^{intra}(\omega)= \frac{2i e^2 k_B T}{\pi \hbar (\omega+i\gamma_c)} \mbox{ln}\left[2 \mbox{cosh}(\mu_c/2 k_B T)\right],
\end{equation}  
\begin{eqnarray} \label{inter}
\sigma^{inter}(\omega)= \frac{e^2}{\hbar} \Bigg\{   \frac{1}{2}+\frac{1}{\pi}\mbox{arctan}\left[(\omega-2\mu_c)/2k_BT\right]-\nonumber \\
   \frac{i}{2\pi}\mbox{ln}\left[\frac{(\omega+2\mu_c)^2}{(\omega-2\mu_c)^2+(2k_BT)^2}\right] \Bigg\},
\end{eqnarray}  
where $\mu_c$ is the chemical potential (controlled with the help of a gate voltage), $\gamma_c$ the carriers scattering rate, $e$ the electron charge, $k_B$ the Boltzmann constant and $\hbar$ the reduced Planck constant.
  
%
  
In all the examples the dielectric elliptical wires are non magnetic ($\mu_2=\mu_3=1$) and $\varepsilon_2 =\varepsilon_3= 3.9$. The graphene parameters are $T = 300 K$ and $\gamma_c = 0.1$ meV.

\subsection{Single wire system: Radiation properties of one emitter coupled to an elliptical graphene wire}

\begin{figure}
\centering
\resizebox{0.5\textwidth}{!}
{\includegraphics{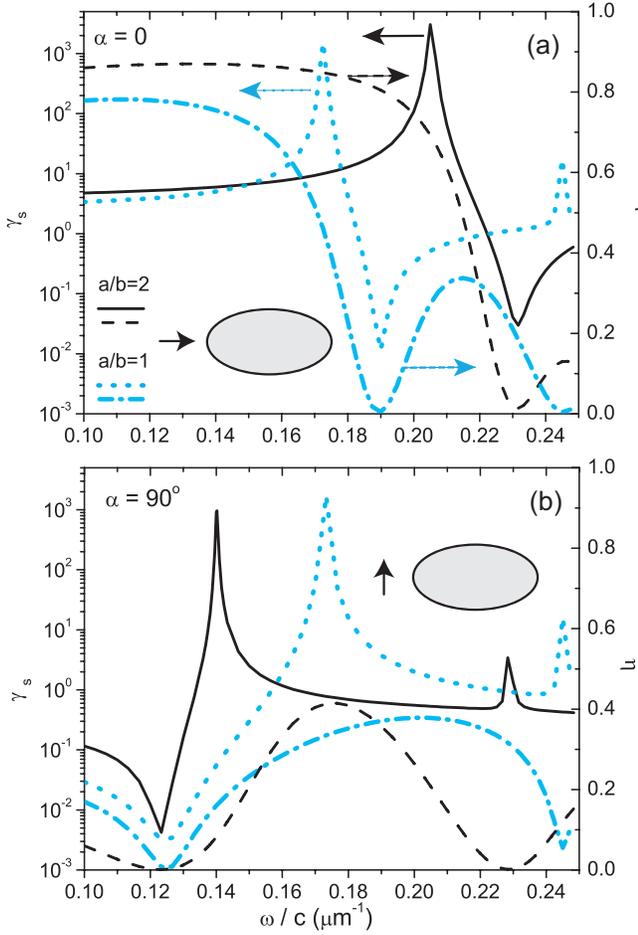}}
\caption{\label{fig:epsart} Radiative decay rate and quantum efficiency, both as a function of frequency, of a graphene coated elliptical wire with a major to minor semi--axes ratio $a/b=2$ ($a\approx 0.65,\,b\approx 0.325$).  Constitutive parameters  $\varepsilon_2=3.9$, $\mu_2=1$ and $\varepsilon_1=\mu_1=1$. The graphene parameters are $\mu_c=0.5$eV, $T = 300$ K and  $\gamma_c=0.1$meV. The emitter is localized on the $\hat{x}$ axis (major ellipse semi--axis) at a distance $d=0.1\mu$m from the left corner of the ellipse. (a) Horizontally orientation, (b) vertically orientation. The light cyan curve corresponds to a graphene--coated  circular wire
($a = b = 0.5 \mu$m).  
}
\label{1elipse}
\end{figure}

Firstly, we use GSIM given by Eq. (\ref{s1wire}) to examine  the elementary system of an emitter coupled to a graphene wire of elliptical cross section. In Figure \ref{1elipse} we plot the radiative decay rate $\gamma_s$ and the quantum efficiency $\eta$ for an emitter located on the $\hat{x}$ axis at $d=0.1\mu$m from the  corner  of an elongated graphene wire whose length perimeter is equal to $3.14 \mu$m  and  with a  major (along $\hat{x}$ axis) to minor (along $\hat{y}$ axis) semi--axes ratio $a/b = 2$. To illustrate the  effects of varying the $\alpha$ orientation angle, in Figure \ref{1elipse}a and \ref{1elipse}b we have plotted these curves for $\alpha=0$ (horizontal polarization) and for $\alpha=90^\circ$ (vertical polarization), respectively. The curves corresponding to the circular wire $a=b$ of the same perimeter is given as a reference.   We observe that, similar to the case of quasi--rectangular graphene wires \cite{cuevas3}, the break of the $90^\circ$ rotational symmetry introduce an anisotropy in the optical behavior. This anisotropy is evident for the dipolar plasmonic  resonance which for the case of circular cross section occur near $0.17\mu$m$^{-1}$ and that is split into two peaks, one of them near $0.14\mu$m$^{-1}$ and the other near $0.205\mu$m$^{-1}$. The first peak corresponds to $\alpha=90^\circ$  
while the second peak corresponds to $\alpha=0$, 
as clearly indicated in Figure \ref{1elipse}  by the fact that both resonances are decoupled for dipole moment orientations parallel  to either of the ellipse's axes and that the first (respectively second) peak is absent when the dipole moment direction is parallel to (respectively along) the major axis.

Another quantity strongly depending on the dipole moment orientation is the quantum efficiency $\eta$.   
Figure \ref{1elipse}a shows that $\eta$ takes a value near $0.7$ at the high plasmonic resonance frequency ($\omega/c\approx0.205\mu$m$^{-1}$), a value considerably greater than that corresponding to  the circular case for which $\eta \approx 0.4$. On the contrary, Figure \ref{1elipse}b shows an insignificant value, near $0.14$, at the low plasmonic resonance frequency.  

On the other hand, we observe a pronounced dip in the radiation efficiency curves above the resonance frequency  for horizontal  polarization (Figure \ref{1elipse}a)  and bellow the resonance frequency for vertical polarization (Figure \ref{1elipse}b). As pointed out in \cite{cuevas4}, these dips result from the destructive interference between the source dipole and that induced in the graphene wire. The minimum value reached at dip positions  are more or less pronounced depending  on whether $\alpha=0$ or $\alpha=90^\circ$, as clearly indicated in Figure \ref{1elipse} by the fact that the minimum, which in the case of $a=b$ and for horizontal polarization occur at $0.19\mu$m$^{-1}$,  is blue shifted to $0.231\mu$m$^{-1}$ in the case of $a=2b$ whereas Figure \ref{1elipse}b shows that  the spectral position of the minimum  for vertical polarization reaches a value of approximately $0.124\mu$m$^{-1}$ regardless whether $a=b$ or $a=2b$.     

\subsection{Dimer Micro--antenna: Radiation properties of a system composed by two graphene coated elliptical wires}

Having studied the radiation properties of a single elliptical  wire, we next explore the effects  that  the incorporation of another wire, to  form a dimmer antenna consisting of two identical elements, has on the emission and the radiation spectrum of a dipole emitter in the proximity of the antenna. In particular, we consider the emitter located  at the gap center of the dimer micro--antenna (see Figure 1).

\begin{figure}
\centering
\resizebox{0.5\textwidth}{!}
{\includegraphics{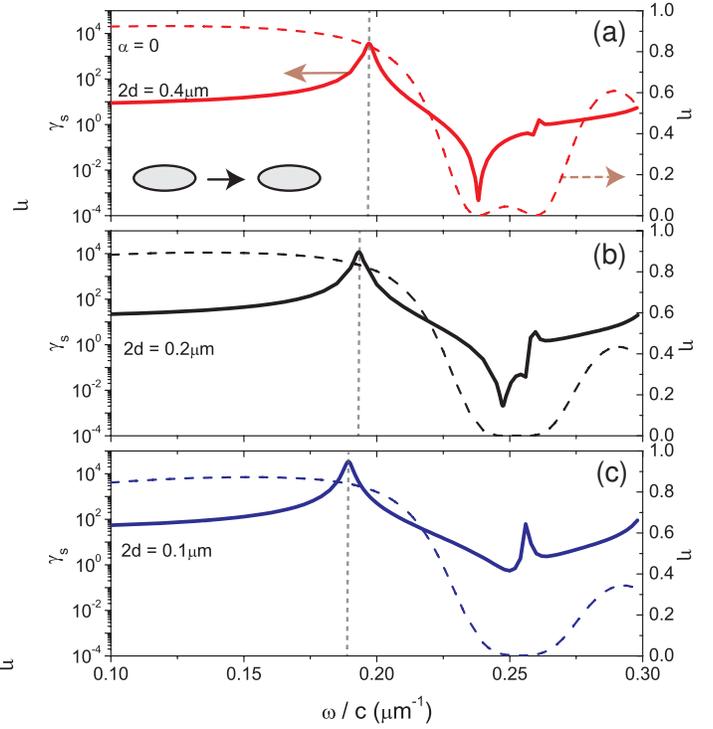}}
\caption{\label{fig:epsart} Radiative decay rate (continuous line) and quantum efficiency (dashed line)  of a micro--antenna composed by two  graphene coated elliptical wires ($a/b=2$,  $a=0.325$). 
The emitter is localized on the  $\hat{x}$ axis (major ellipse semi--axis) at the center of the gap and the orientation angle $\alpha=0$ (horizontal orientation).  Gap size $2d=0.4\mu$m (a), $2d=0.2\mu$m  (b) and $2d=0.1\mu$m (c). The other parameters as the same as in Figure \ref{1elipse}} \label{2elipses}
\end{figure}

Figure \ref{2elipses} shows the normalized radiative decay rate $\gamma_s$ for an emitter horizontally oriented, \textit{i.e.} the dipole moment of the source is oriented along the line connecting both elements, for gap  sizes (the gap between the two wire elements) $2d=0.4\mu$m, $0.2\mu$m and $0.1\mu$m. On comparing Figures \ref{2elipses}a--c, we observe that the high frequency dipolar maximum which in the single wire case occur at $0.205\mu$m$^{-1}$ is red shifted from $0.197\mu$m$^{-1}$ to $0.189\mu$m$^{-1}$ when the gap $2d$ is decreased from $0.4\mu$m to $0.1\mu$m. A similar behavior was found in case of metallic nano--antennas where the dipolar plasmonic resonance of the dimer structure is shifted to longer wavelengths as the gap between the two elements is decreased \cite{rogobete2,sanchezgil}. Figure \ref{2elipses} also shows that the maximum value in the $\gamma_s$ curve at the dipolar resonance is increased when the gap $2d$ is decreased.    
Moreover, the quantum efficiency $\eta$ at the resonant frequency which in the single wire case is approximately $0.7$, reach values slightly higher than $0.8$ for a dimer graphene wire antenna. In particular, the calculated values in Figure \ref{2elipses} are $\eta=0.82,\,0.836\,\mbox{and}\,0.847$ for $2d=4\mu$m, $2\mu$m, and $1\mu$m respectively. 

On the other hand, the non--radiating effect resulting from the destructive interference between the emitter and the antenna is becoming less noticeable as the gap decreases, as clearly indicated in Figure \ref{2elipses} by the fact that the minimum in the $\gamma_s$ curve reaches a value varying from $3\,10^{-4}$ for $2d=0.4\mu$m to  $0.55$ for $2d=0.1\mu$m.

\begin{figure}
\centering
\resizebox{0.5\textwidth}{!}
{\includegraphics{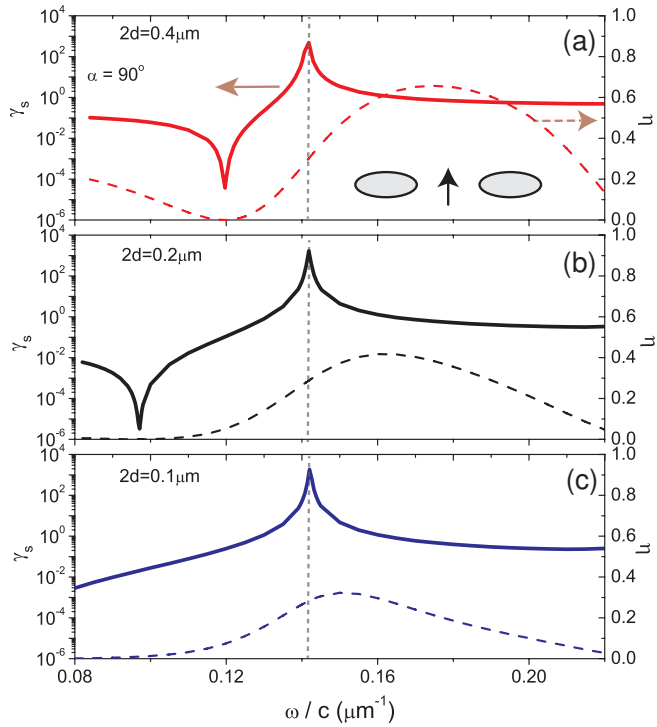}}
\caption{\label{fig:epsart} Radiative decay rate and quantum efficiency  of a micro--antenna composed by two  graphene coated elliptical wires ($a/b=2$,  $a=0.325$). 
The emitter is localized on the  $\hat{x}$ axis (major ellipse semi--axis) at the center of the gap and the orientation angle $\alpha=90$ (vertical orientation).  Gap size $2d=0.4\mu$m (a), $2d=0.2\mu$m  (b) and $2d=0.1\mu$m (c). The other parameters as the same as in Figure \ref{1elipse}} \label{2elipses90}
\end{figure}

In  Figure \ref{2elipses90} we plot the normalized radiative decay rate $\gamma_s$ for an emitter vertically oriented, \textit{i.e.}, with its dipole moment perpendicular to the connection line of both  wires, for the same gap size  values as in Figure \ref{2elipses}. Unlike the horizontally orientation case in which the spectral position of the resonance peak is rather dependent on the gap size, in Figure \ref{2elipses90} we see that the frequency at which the low frequency dipolar maximum occur almost does not shows any dependence on the gap size. We also see that the quantum efficiency $\eta$ at the resonance frequency reaches values close to $0.3$, a negligible value  when it is compared with the values of $\eta$ obtained in the horizontally orientation case.

An interesting result obtained in the vertical orientation is the strong reduction of the radiation decay rate at frequencies where non--radiating states occur, as clearly indicated in Figure \ref{2elipses90} by the fact that the low frequency minimum in the $\gamma_s$ curve is more pronounced as the gap size is decreased. For instance, $\gamma_s$ takes a minimum value $\approx 5\,10^{-5}$ for a  gap size $2d=4\mu$m  (Figure \ref{2elipses90}a), a value $\approx 2\,10^{-6}$ for $2d=2\mu$m  (Figure \ref{2elipses90}b) and a value $\approx \,10^{-7}$ for $2d=1\mu$m  (this last does not shown in Figure \ref{2elipses90}c). Moreover, on comparing Figures \ref{2elipses90}a--c we see that the minimum position is red shifted as the gap size is decreased. This behavior can be understood  by taking into account that in the single wire case the spectral position of the non--radiating states are red shifted as the distance between the emitter, vertically oriented,   and  the wire is decreased \cite{cuevas4}.

\begin{figure}
\centering
\resizebox{0.50\textwidth}{!}
{\includegraphics{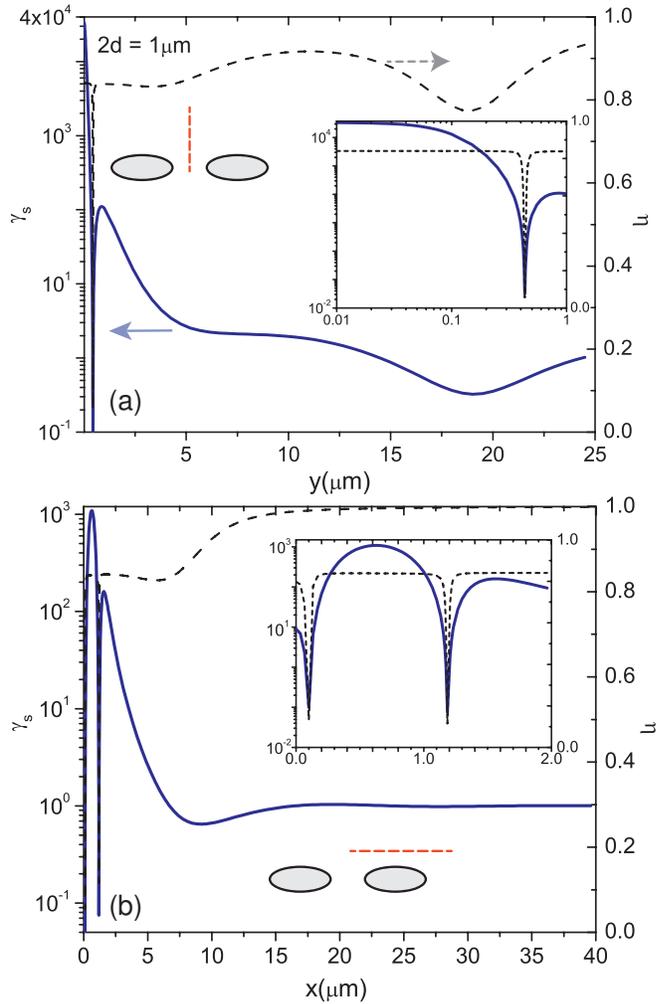}}
\caption{\label{fig:epsart} Radiative decay rate and quantum efficiency as a function of the emitter location  for the dipole resonance frequency.  In (a) the dipole trajectory is along the $\hat{y}$ axis (see inset), and in (b) the dipole trajectory is along the $\hat{x}$ axis (see inset). The dipole moment is oriented along $\hat{x}$ axis (horizontal orientation) and the dimer gap size is $2d=0.1\mu$m. All other the parameters are the same as in Figure \ref{1elipse}. 
} \label{posicion}
\end{figure}
To evaluate the dependence of the radiation properties with the location of the emitter, in Figure \ref{posicion} we plot the radiation decay rate $\gamma_s$ and the quantum efficiency $\eta$ for an emitter as it   
moves away from the gap center of the micro--antenna (gap size $2d=1\mu$m).  
Since the quantum efficiency for vertically polarization is low, we only exemplify the case of  horizontal polarization for which the achieved efficiency is $\approx 80$\%. The emission frequency is chosen as the high dipolar plasmonic resonance $\omega/c=0.18934\mu$m$^{-1}$  in Figure \ref{2elipses}c.  
  
Firstly, the emitter is displaced vertically from the gap center as indicated in the inset in Figure \ref{posicion}a. We observe that the decay rate curve strongly decreases from its maximum value at the gap center  to its minimum value from which the curve  increases leading to an oscillatory behavior whose period $\approx \lambda/2=\pi/(\omega/c)=16.5\mu$m. This oscillatory behavior which occur for $y>5\mu$m also found in case of metallic dimer nano--antennas \cite{sanchezgil}. 
 To appreciate the details at locations near the micro--antenna, in the inset of Figure \ref{posicion}a we have enlarged the horizontal scale, where it is clearly shown that a pronounced minimum is reached at a position slightly higher that the end of the elliptical wire composing the micro--antenna. 
In addition, the quantum efficiency takes a value $\approx 0.8$ for $y$ values that are lower than $5\mu$m except in the neighborhood of $y=0.43\mu$m where the $\eta$ curve reaches a minimum value $\approx 3\,10^{-2}$.
  For $y>5\mu$m, this  curve presents the same oscillatory behavior as the $\gamma_s$ curve.   

We next consider that the emitter, with horizontally orientation, is displaced horizontally from the position placed at $x=0,\,y=0.4\mu$m, as is indicated in the inset of Figure  \ref{posicion}b.  We see that the radiative decay rate decreases from the maximum value, $\approx 10$, at the gap center to a sharp minimum at approximately the position of the ellipse corner facing the gap (inset in Figure \ref{posicion}b). Next, the $\gamma_s$ curve increases reaching a maximum value at the center of the elliptical wire and then it  decreases to another sharp minimum at the position of the far ellipse corner. On the other hand, the quantum efficiency takes a value $\approx 0.8$ reaching sharp minima values coinciding with that of the decay radiative curve. For $x$ values larger than $15\mu$m both the radiative decay rate and the quantum efficiency take a value close to unity, suggesting that the emitter is uncoupled from the micro--antenna.   
\begin{figure}
\centering
\resizebox{0.48\textwidth}{!}
{\includegraphics{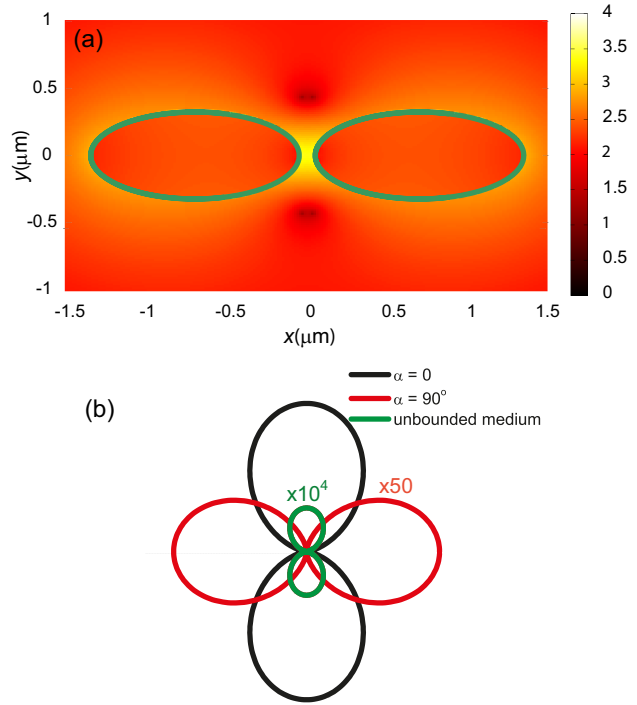}}
\caption{\label{fig:epsart} (a) Map of the scattered electric field modulus in logarithmic scale at a fixed time for a horizontal dipole at the gap center of the micro--antenna. 
(b) Far field intensities for the dipolar plasmon resonance of a horizontal (black line) and a vertical (red line) dipole placed at the gap center. For comparison, the isolated dipole contribution is also shown for the horizontal  polarization (green line).  The frequencies $\omega/c = 0.18934 \mu$m$^{-1}$ (horizontal dipole) and     $\omega/c = 0.1421 \mu$m$^{-1}$ (vertical dipole) coincide with those of the dipolar plasmon resonances.
The gap size of the micro--antenna $2d=0.1\mu$m and all other parameters are the same as in Figure \ref{2elipses}.} \label{campos}
\end{figure}

In Figure \ref{campos}a we plot the spatial distribution of the scattered near electric field for an emitter polarized along the $\hat{x}$ axis ($\alpha=0$) and placed at the  gap center of the micro--antenna ($2d=1\mu$m). The  chosen emission frequency corresponds to the high dipolar resonance $\omega/c=0.189\mu$m$^{-1}$.   We can see a strongly enhancement of the field in the gap region and, to a lesser extent, around each wire.  
We have obtained a similar spatial field distribution for an emitter polarized along the $\hat{y}$ axis (not shown in Figure \ref{campos}) but with a field enhancement that is near one order of magnitude less than the corresponding to the horizontal polarization. The strongly confinement of the field in the gap region, where the emitter is located, is responsible of the emitted power enhancement observed in Figures \ref{2elipses}c and \ref{2elipses90}c at a frequency $\omega/c=0.189\mu$m$^{-1}$ and $\omega/c=0.142\mu$m$^{-1}$ respectively. However, the two configurations noticeably differ in the far field emission, as can be seen in Figure \ref{campos}b where we have plotted the far field intensity as a function of the angle of radiation. We see that the far field intensity for the horizontal polarization is markedly greater (near $50$ times) than that corresponding to the vertical polarization. This can be understood by taking into account that the quantum efficiency reach a value near to $0.85$ for the horizontal polarization and a value near to $0.2$ for the vertical polarization and that the field enhancement in the gap region for the horizontal polarization is near $10$ times greater than that corresponding to the vertical  polarization.     

\begin{figure}
\centering
\resizebox{0.48\textwidth}{!}
{\includegraphics{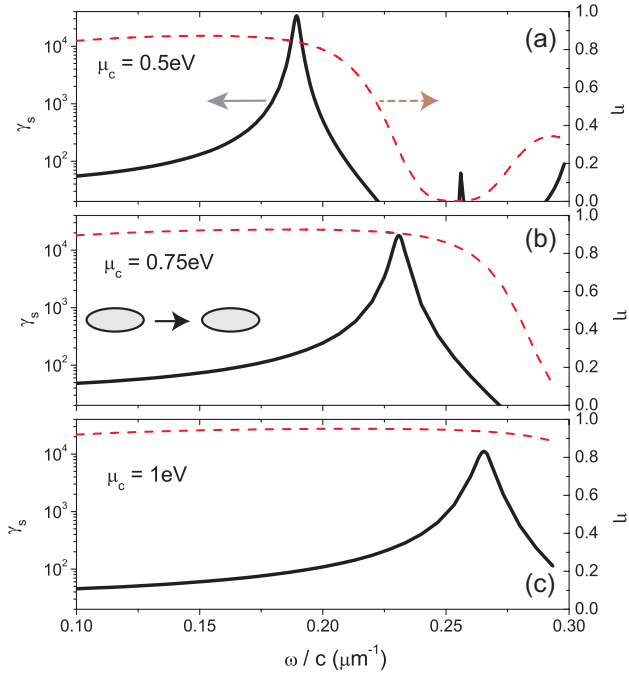}}
\caption{\label{fig:epsart} Radiative decay rate and quantum efficiency as a function of frequency for different values of the chemical potential $\mu_c$, (a) $\mu_c=0.5$eV, (b) $\mu_c=0.75$eV, (c) $\mu_c=1$eV. All other the parameters are the same as in Figure \ref{2elipses}.} \label{mu}
\end{figure}

\begin{figure}
\centering
\resizebox{0.48\textwidth}{!}
{\includegraphics{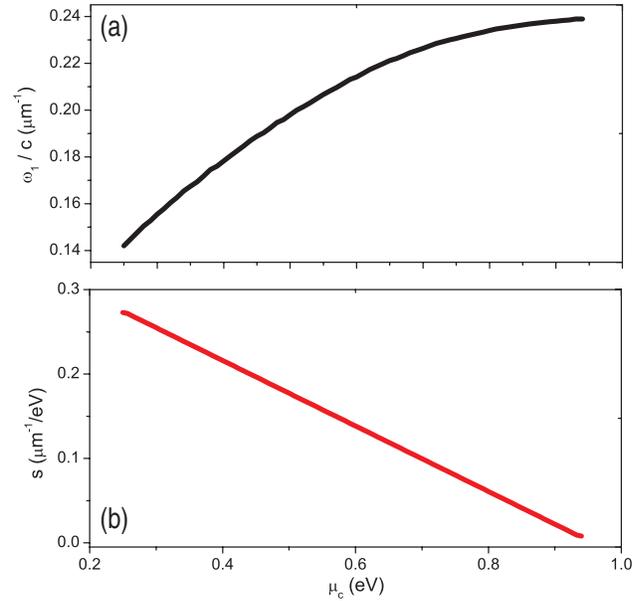}}
\caption{\label{fig:epsart} Resonance frequency (a) and sensitivity $s$ (b) as a function of the chemical potential $\mu_c$ for the high dipolar resonance (the resonance corresponding to the horizontal polarization).  All other the parameters are the same as in Figure \ref{2elipses}.} \label{s}
\end{figure}

Finally, we investigate the tunability of the graphene micro--antenna by varying the chemical potential of the graphene sheets. We focused on the symmetric case where two wires are tuning at the same $\mu_c$. 
 By increasing the chemical potential, we are able to tune the resonance frequency and we can define the sensitivity $s$ to the chemical potential as,
\begin{equation}
s=\frac{\partial \omega / c}{\partial \mu_c}.  
\end{equation}
In Figure \ref{mu} we plotted the frequency dependence of the radiative decay rate and the quantum efficiency for $\mu_c=0.5,\,0.75,\,\mbox{and}\,1$eV. The emitter is localized at the gap center and with its dipole moment along the $\hat{x}$ direction. On comparing Figures \ref{mu}a--c, we observe that the dipolar resonance peak is blue shifted from $0.1893\mu$m$^{-1}$ to $0.2654\mu$m$^{-1}$ when the chemical potential $\mu_c$ is increased from $0.5$eV to $1$eV. This fact can be understood by the fact that an increase in the chemical potential leads to an increase in the surface charge density on graphene sheets and, as a consequence, the plasmon resonance frequency increases.  
Figures \ref{s}a and b show the resonance frequency and the sensitivity $s$  as a function of the chemical potential $\mu_c$.  The values of the resonance frequencies have been indirectly calculated  estimating from the observation the positions of maxima of resonances in radiative decay rate curve spectra obtained for different values of $\mu_c$. We observe that the resonance  frequency quadratically increases with the chemical potential. As a consequence, and as it can be seen in Figure \ref{s}b, the sensitivity $s$ linearly  decreases with the chemical potential increase.
Another interesting results is the increment of the quantum efficiency at the dipolar resonance frequency with the chemical potential. For example, $\eta \approx 0.84$ for $\mu_c=0.5$eV whereas $\eta$ reaches values  $\approx 0.91$ and $\approx 0.94$   for $\mu_c=0.75$eV and $\mu=1$eV, respectively.

\section{Conclusions} \label{conclusiones}

The emission and radiation properties of a dipole emitter source close to a dimer graphene plasmonic antenna have been studied by applying an electromagnetically rigorous integral  method based on the Green second identity. We considered the case of graphene--coated wires of elliptical section. In comparison with the  circular section case, and as it might be expected on symmetry grounds, a frequency splitting of the dipolar plasmonic resonance is observed in the emission and  radiation decay rate spectra. The high dipolar resonance  (the resonance corresponding to the horizontal polarization)  is  red shifted as the gap size between the two  wire components is decreases, whereas the spectral position of the low dipolar resonance (the resonance   corresponding to the vertical polarization)  is almost independent from the gap size value. 

Our results shown similar values of the Purcell factor for both polarizations. For instance, the emission decay rate takes a value near $10^4$ for the horizontal polarization and a value near $10^3$ for the vertical polarization. 
However, the density of radiating states for horizontal polarization is larger than that corresponding to the vertical polarization, as indicated by calculated quantum efficiency values. For instance, the quantum efficiency  is near $0.8$ for horizontal polarization and near $0.3$ for vertical polarization. This result means that a desirable antenna feature is achieved when the dipole moment of the emitter is oriented along the line connecting both elliptic wires. 
Moreover, we have calculated the far field intensity which have served as a further test to confirm the previous guess. 

Another interesting result revealed in this study is the fact that the quantum efficiency is strongly depending of the 
chemical potential of graphene coating. By increasing the chemical potential, the quantum efficiency is notably increased leding to an improve in the antenna performance.




\section*{Acknowledgment}
The author acknowledge the financial support of Consejo Nacional de Investigaciones Cient\'{\i}ficas y T\'ecnicas, (CONICET, PIP 451).

\end{document}